\documentclass[10pt, conference, compsocconf]{IEEEtran}
\usepackage{balance}
\usepackage[cmex10]{amsmath}
\usepackage{amssymb}
\usepackage{url}
\usepackage[noadjust]{cite}
\usepackage{color}
\usepackage{array,booktabs}
\usepackage[usenames,dvipsnames]{xcolor}

\usepackage{algorithm}
\usepackage{algorithmicx}
\usepackage[noend]{algpseudocode}

\usepackage{mdwmath}
\usepackage{textcomp}
\usepackage{xspace}
\usepackage{graphicx}

\usepackage{enumitem}
\usepackage[scaled=.7]{beramono}
\usepackage{etoolbox,siunitx}
\robustify\bfseries
\newcommand{\rb}[1]{\raisebox{1.5ex}[0pt]{#1}}
\newcommand{\comment}[1]{}

\def\IEEEbibitemsep{0pt plus .5pt}
\newcommand\verysmallfont{\fontsize{8}{9}\selectfont}
\newcommand\supersmallfont{\fontsize{7}{8}\selectfont}

\newcommand\newtilde{\raise.17ex\hbox{$\scriptstyle\sim$}}

\newcommand{\Pulp}{\textsc{PuLP}\xspace}
\newcommand{\Xtrapulp}{\textsc{XtraPuLP}\xspace}
\newcommand{\Hyperpulp}{\textsc{Hyper-Pulp}\xspace}

\algnotext{EndFor}
\algnotext{EndIf}
\algnotext{EndWhile}
\algnotext{EndProcedure}

\begin{document}

\title{Partitioning Trillion-edge Graphs in Minutes}

\author{
\IEEEauthorblockN{George M. Slota}
\IEEEauthorblockA{Computer Science Department\\
Rensselaer Polytechnic Institute\\
Troy, NY\\
slotag@rpi.edu}
\and
\IEEEauthorblockN{Sivasankaran Rajamanickam \& Karen Devine}
\IEEEauthorblockA{Scalable Algorithms Department\\
Sandia National Laboratories\\
Albuquerque, NM\\
srajama@sandia.gov}
\and
\IEEEauthorblockN{Kamesh Madduri}
\IEEEauthorblockA{Computer Science and Engineering\\
The Pennsylvania State University\\
University Park, PA\\
madduri@cse.psu.edu}
}

\maketitle

\begin{abstract}
We introduce \Xtrapulp, a new distributed-memory graph partitioner designed to process trillion-edge graphs. \Xtrapulp is based on the scalable label propagation community detection technique, which has been demonstrated as a viable means to produce high quality partitions with minimal computation time. On a collection of large sparse graphs, we show that \Xtrapulp partitioning quality is comparable to state-of-the-art partitioning methods. We also demonstrate that \Xtrapulp can produce partitions of real-world graphs with billion+ vertices in minutes. Further, we show that using \Xtrapulp partitions for distributed-memory graph analytics leads to significant end-to-end execution time reduction.
\end{abstract}


\section{Introduction}
\label{s:intro}


We introduce \Xtrapulp, a new graph partitioner exploiting distributed-memory parallelism plus threading to efficiently partition extreme-scale real-world graphs. \Xtrapulp can be considered a significant extension to our prior shared-memory-only partitioner, \Pulp~\cite{pulp}. Graph partitioning is an essential preprocessing step to ensure load-balanced computation and to reduce inter-node communication in parallel applications~\cite{part_survey, PothenOldSurvey}. With the sizes of online social networks, web graphs, and other non-traditional graph data sets (e.g.\ brain graphs) growing at an exponential pace, scalable and efficient algorithms are necessary to partition and analyze them. Online social networks and web crawls are typically characterized by highly skewed vertex degree distributions and low average path lengths. Some of these graphs can be modeled using the ``small-world'' graph model~\cite{watts1998collective, kleinberg2000small}, and others are referred to as ``power-law'' graphs~\cite{faloutsos1999power, barabasi1999emergence}.

For highly parallel distributed-memory graph analytics on billion+ vertex or trillion+ edge small-world graphs, any computational and communication imbalance can result in a significant performance loss. Thus, graph partitioning can be used to improve balance. Traditional methods for partitioning graphs (e.g.\ ParMETIS) are limited either in the size of the graphs they can partition, or the partitioning objective metrics that they can support. Furthermore, as the analytics themselves are often quite fast in comparison to other scientific computing applications, the time and scalability requirements for the effective use of a graph partitioner with these applications is much stricter.

In essence, partitioning methods targeting emerging graph analytics should be significantly faster than the current state-of-the-art, support multiple objective metrics, scale better on irregularly structured inputs, and scale to emerging real-world problem sizes. We also desire the method to (1) be more memory-efficient than partitioning methods used for traditional scientific computing problems; (2) have good strong-scaling performance, since we may work with fixed-size problems; (3) be relatively simple to implement, and (4) require little tuning.

There has been some progress made in the recent past towards such partitioning methods. There are methods that are based on random or edge-based distributions~\cite{yoo}, label propagation-based graph partitioning methods~\cite{partbillion, UG13}, adaptations of traditional partitioning methodologies to small-world graph instances~\cite{kahipPar}, and methods using two-dimensional distributions~\cite{2D}. Among these, label propagation-based techniques are the most promising in terms of meeting our requirements. We consider extending these techniques for graph inputs several orders-of-magnitude larger than previously processed by any other partitioner. Problems at the trillion edge scale have been attempted only recently~\cite{ching,checconi}, but not in the context of a problem as computationally challenging as graph partitioning.



The following are our key contributions:
\begin{itemize}[leftmargin=*]
  \item We describe \Xtrapulp, a distributed-memory partitioning method that can scale to graphs with billion+ vertices and trillion+ edges. Implementing a partitioning algorithm at this scale (relative to e.g.\ a billion edges, the limit of traditional methods) requires careful consideration to computation, communication, and memory requirements of the partitioner itself. Significant changes from our shared-memory partitioner \Pulp are required, including development of entirely new routines for in-memory graph storage, inter-node communication, and processing of part assignment updates.
        
  \item We demonstrate the scalability of our MPI+OpenMP parallel partitioner by running on up to 131,072 cores of the NCSA Blue Waters supercomputer, using graph instances with up to 17 billion vertices and 1.1 trillion edges.

  \item We demonstrate state-of-the-art partitioning quality for computing partitions satisfying multiple constraints and optimizing for multiple objectives simultaneously. We show comparable quality relative to \Pulp and ParMETIS.

  \item We utilize partitions from \Xtrapulp in two settings. First, we demonstrate reduction in end-to-end time for six  graph analytics with various performance characteristics. Second, we show reduction in time for parallel sparse matrix vector multiplications with two dimensional matrix layouts calculated from \Xtrapulp's vertex partitions.

\end{itemize}






\section{Background}
\label{s:background}

\subsection{Graph Partitioning}

Given an undirected graph $G=(V,E)$ and vertex and edge imbalance ratios $\mathit{Rat}_v$ and $\mathit{Rat}_e$ and target max part sizes $\mathit{Imb}_v$ and $\mathit{Imb}_e$, the graph partitioning problem can be formally described as partitioning $V$ into $p$ disjoint parts. Let $\Pi = \left\{ \pi_1, \ldots, \pi_{p} \right\}$ be a balanced partition such that $\forall~i=1\ldots p$,
\begin{eqnarray}
   |V(\pi_i)| &\le \left(1+ \mathit{Rat}_v \right) \dfrac{|V|}{p} &= \mathit{Imb}_v \\
   |E(\pi_i)| &\le \left(1+ \mathit{Rat}_e \right) \dfrac{|E|}{p} &= \mathit{Imb}_e
\end{eqnarray}

$V(\pi_i)$ is the set of vertices in part $\pi_i$ and $E(\pi_i)$ is the set of edges such that both its endpoints are in part $\pi_i$. We define the set of cut edges as

  $C(G,\Pi) = \left\{ \{(u,v) \in E \}~|~\Pi(u) \ne \Pi(v) \right\}$,

\noindent and set of cut edges in any part as

  $C(G,\pi_k) = \left\{ \{(u,v) \in C(G,\Pi) \}~|~(u \in \pi_k \lor v \in \pi_k) \right\}$. 

\noindent Our partitioning problem is then to minimize the two metrics
  $|C(G, \Pi)|$ and $\max_{k} |C(G, \pi_k)|$.

\subsection{Label Propagation}

The label propagation community detection algorithm~\cite{labelprop} is a fast and scalable method for detecting communities in large networks. The primary motivation for using label propagation for the community detection problem is that its per-iteration cost is linear in the graph size, or $O(|V|+|E|)$. Label propagation is also shown to find communities in a constant number of iterations on several real-world graphs. Community detection methods also naturally lend themselves towards use in partitioning, as the optimization problems solved are similar. Community detection algorithms attempt to find tightly knit communities having a high relative portion of internal edges among members of the community versus external edges to members of other communities. This goal corresponds to a partitioning method attempting to separate a graph into some number of parts where each part has a high number of internal and few external (cut) edges. 

\subsection{Related Work}

Due to its linear work bound, scalability, and similar optimization goal,
label propagation has seen relatively widespread adoption as an effective means to find high quality partitions of small-world and irregular networks, such as social networks and web crawls. There are two primary approaches for using label propagation in partitioners. 

The first approach uses label propagation as part of a multilevel framework. In multilevel partitioning, an input graph is iteratively coarsened to a much smaller graph, the coarse graph is partitioned, and then iterative stages of uncoarsening and partition refinement take place until a partition of the original graph is produced. Partitioners that utilize these techniques include Meyerhenke et al.~\cite{kahipPar} and Wang et al.~\cite{partbillion}. Wang et al. demonstrated a case study of how label propagation might be used as part of a multilevel partitioner, by first coarsening the graph in parallel and then running METIS~\cite{METIScode} at the coarsest level. Meyerhenke et al. improved upon this approach in terms of partition quality and execution time by running an optimized implementation of distributed label propagation and then parallel runs of the evolutionary algorithm-based state-of-the-art KaFFPaE partitioner at the coarsest level. The biggest drawbacks to these methods, and multilevel methods in general, are the high memory requirements that result from having to store copies of the graph at the multiple levels of coarsening, and the need to use poorly scaling or serial partitioning methods at the coarsest level. Multilevel methods have not been experimentally demonstrated to process irregular graphs larger than approximately $O(1~\mathit{billion})$ edges in size.

The second approach uses label propagation directly to compute the partitions. Early efforts utilizing this approach include Ugander et al.~\cite{UG13} and Vaquero et al.~\cite{xdgp}. Wang et al.~\cite{partbillion} additionally used a variant of their coarsening scheme to compute balanced partitions, although at a non-negligible cost to cut quality. In general, this cost was observed in early single level methods, which demonstrated good scalability and performance, but often with a high cost in terms of partition quality. In our recent prior work, we introduced \Pulp~\cite{pulp}, which uses weighted label propagation variants for various stages of a multi-constraint and multi-objective shared memory parallel partitioning algorithm. Buurlage~\cite{hyperpulp} extended our initial work with \Hyperpulp, which modified the general \Pulp scheme to the distributed partitioning of hypergraphs. Note that hypergraph partitioning requires a significantly different approach than graph partitioning. We only perform graph partitioning in our work due to considerably lower overheads and higher scalability relative to hypergraph partitioning. The graphs we're considering in our work are over 20 million times larger than those partitioned with \Hyperpulp.

Our work extends these two recent efforts significantly, as we strive to offer a highly performant label propagation-based distributed parallel partitioner that also computes high quality partitions of very large, irregular input graphs.

\section{\Xtrapulp}
\label{s:algorithm}


This section provides algorithmic and implementation details of \Xtrapulp, our distributed-memory label propagation-based partitioner. We note explicitly that our primary contribution is technical and not algorithmic, in that we provide a discussion of the technical necessities to scale the prior \Pulp algorithms to process graphs of several orders-of-magnitude larger and on several orders-of-magnitude more cores than the prior implementation is capable. Three main extensions needed for the distributed implementation relative to \Pulp are:
\begin{itemize}
\item The graph and its vertices' part assignments and other associated data must be distributed in a memory-scalable way across processors. Only the necessary local per-task information should be stored to reduce memory overhead. Access to task-specific information should also be as efficient as possible for computational scalability in a large cluster. We develop and optimize our implementation to achieve these objectives.
\item MPI-based communication is needed to update boundary information and  compute the global quantities required by out weighting functions. We implement highly optimized communication routines to achieve scaling to thousands of nodes.
\item The update pattern of part assignments must be finely controlled to prevent wild oscillations of part assignments as processes independently label their vertices. We use a dynamic multiplier $\mathit{mult}$ that iteratively adjusts to enable partitions to attain balance in a more stable fashion. We also analyze the parameters controlling $\mathit{mult}$ and its effect on partition quality and achieving the balance constraints.
\end{itemize}

Additionally, we offer a novel initialization strategy that is observed to substantially improve final partition quality for certain graphs, while not negatively impacting partition quality for other graphs.

\subsection{\Xtrapulp Overview}

\paragraph{\bf{Graph Representation}} We use a distributed one-dimensional compressed sparse row-like representation, where each task owns a subset of vertices and their incident edges (representing a local graph $G(V,E)$). When distributing the graph for the partitioner, we utilize either random and block distributions of the vertices. We observe random distributions are more scalable in practice for irregular networks. Each vertex's global identifier is mapped to a task-specific local one using a hash map. Local to global translation uses values stored in a flat array. Each task stores part labels for both its owned vertices as well as its ghost vertices (vertices in its one hop neighborhood that are owned by another task). When computing the partition, a task will calculate updates only for its owned vertices and communicate the updates so the task's neighbors update assignments for the ghosts. 

\begin{algorithm}[thb]
  \verysmallfont
  \begin{algorithmic}[htb]
    \Procedure{\Xtrapulp}{$G(V, E)$}
    \State $\mathit{parts} \gets$ \Xtrapulp-Init($G(V, E)$)
    \State {$I_{\text{outer}} \gets 3$ \qquad $I_{\text{bal}} \gets 5$ \qquad $I_{\text{ref}} \gets 10$}
    \State $I_{\text{tot}} \gets I_{\text{outer}} \times (I_{\text{bal}} + I_{\text{ref}})$
    \State $\mathit{Iter}_{\text{tot}} \gets 0$
    \State $\mathit{Imb}_v \gets$ targetMaxVerticesPerPart()
    \State $\mathit{Imb}_e \gets$ targetMaxEdgesPerPart()
    \For {$iter = 1 \ldots I_{\text{outer}}$}
      \State \Xtrapulp-VertBalance($G(V, E), \mathit{parts}, I_{\text{bal}}, \mathit{Imb}_v$)
      \State \Xtrapulp-VertRefine($G(V, E), \mathit{parts}, I_{\text{ref}}, \mathit{Imb}_v$)
    \EndFor
    \State $\mathit{Iter}_{\text{tot}} \gets 0$
    \For {$iter = 1 \ldots I_{\text{outer}}$}
      \State \Xtrapulp-EdgeBalance($G(V, E), \mathit{parts}, I_{\text{bal}}, \mathit{Imb}_v, \mathit{Imb}_e$)
      \State \Xtrapulp-EdgeRefine($G(V, E), \mathit{parts}, I_{\text{ref}}, \mathit{Imb}_v, \mathit{Imb}_e$)
    \EndFor
    \State \textbf{return} $\mathit{parts}$
    \EndProcedure
  \end{algorithmic}
  \caption{\Xtrapulp Multi-Constraint Multi-Objective Algorithm}
  \label{alg:pulpmm}
\end{algorithm}

\paragraph{\noindent\bf{\Xtrapulp Algorithm}} 

We implement and optimize the original \Pulp-MM algorithm from~\cite{pulp} for multiple objective (minimizing the global cut and maximal cut edges of any part) and multiple constraint (vertex and edge balance) partitioning. An overview of the three stage algorithm is given in Algorithm~\ref{alg:pulpmm}. The first stage is a fast initialization strategy allowing large imbalance among partitions. The second stage balances the number of vertices for each part while minimizing the global number of cut edges. The final stage balances vertices \emph{and} edges, and minimizes the global edge cut \emph{and} maximal edges cut on any part. We have observed in practice that minimizing the maximal per-part cut has the side affect of also balancing the cut edges among all parts. We alternate between balance and refinement stages for $I_{\text{outer}}$ iterations. Each balance and refinement algorithm iterates $I_{\text{bal}}$ or $I_{\text{ref}}$ times internally. We also track $\mathit{Iter}_{\text{tot}}$ and $I_{\text{tot}}$ as part of our distributed communication strategy (explained below). The default values for $I_{\text{outer}}$, $I_{\text{bal}}$, and $I_{\text{ref}}$ are shown in Algorithm~\ref{alg:pulpmm} and used in all experiments.

\subsection{\Xtrapulp Initialization}

We introduce the \Xtrapulp initialization algorithm (Algorithm~\ref{alg:pulp_init}), a hybrid between the two shared-memory \Pulp initialization strategies of unconstrained label propagation~\cite{pulp} and breadth-first search-based graph growing~\cite{pulp2, METISpaper, graphgrowing}. We utilize a bulk synchronous parallel approach for all the stages, while maximizing intra-task parallelism through threading and minimizing communication load with a queuing strategy for pushing updates among tasks.

\begin{algorithm}[htb]
\verysmallfont
\begin{algorithmic}[htb]
\State $\mathit{procid} \gets$ localTaskNum()
\If{$\mathit{procid} = 0$}
  \State $Roots(1 \ldots p) \gets$ UniqueRand($1 \ldots |V_{\mathit{global}}|$)
\EndIf
\State Bcast($Roots$)
\State $\mathit{parts}(1 \ldots |V|) \gets -1$
\For{$i = 1 \ldots p$}
  \If{$Roots(i) \in V$}
    \State $\mathit{parts}(Roots(i)) \gets i$
  \EndIf
\EndFor
\State $updates \gets p$
\While{$updates > 0$}
  \State $updates \gets 0$
  \ForAll{$v \in V$} \Comment{across threads}
    \If{$parts(v) = -1$}
      \State $isAssigned(1 \ldots p) \gets \textbf{false}$
      \ForAll{$\langle v, u\rangle \in E$}
        \If{$\mathit{parts}(u) \ne -1$}
          \State $isAssigned(\mathit{parts}(u)) \gets \textbf{true}$
          \State $updates \gets updates + 1$
        \EndIf
      \EndFor
      \State $w \gets$ RandTrueIndex($isAssigned$)
      \If{$w \neq -1$}
        \State $Q_{\mathit{thread}} \gets \langle v, w\rangle$
      \EndIf
    \EndIf
  \EndFor
  \State $Q_{task} \gets Q_{\mathit{thread}}$ \Comment{merge thread into task queue}
  \State $Q_{recv} \gets$ ExchangeUpdates($\mathit{parts}$, $Q_{task}$, $G$)
  \ForAll{$\langle v, w\rangle \in Q_{recv}$} \Comment{across threads}
    \State $\mathit{parts}(v) \gets w$
  \EndFor
\EndWhile
\ForAll{$v \in V$} \Comment{across threads}
  \If{$\mathit{parts}(v) = -1$}
    \State $\mathit{parts}(v) \gets$ Rand($1 \ldots p$)
    \State $Q_{\mathit{thread}} \gets \langle v, \mathit{parts}(v)\rangle$
  \EndIf
\EndFor
\State $Q_{task} \gets Q_{\mathit{thread}}$ \Comment{merge thread into task queue}
\State $Q_{recv} \gets$ ExchangeUpdates($G$, $\mathit{parts}$, $Q_{task}$)
\ForAll{$\langle v, w\rangle \in Q_{recv}$} \Comment{across threads}
  \State $\mathit{parts}(v) \gets w$
\EndFor
\end{algorithmic}
\caption{\Xtrapulp Initialization: \newline $\mathit{parts} \gets$ \Xtrapulp-Init($G(V,E)$)}
\label{alg:pulp_init}
\end{algorithm}

The master task (process 0) first randomly selects $p$ unique vertices from the global vertex set (array $Roots$) and broadcasts the list to other tasks. Each task initializes its local part assignments to $-1$, and then, if it owns one of the roots, assigns to that root a part corresponding to the order in which that root was randomly selected.

In each iteration of the primary loop of the initialization algorithm, every task considers all of its local vertices that are yet to be assigned a part using thread level parallelism. For a given unassigned local vertex $v$, all neighbors' part assignments (if any) are examined. Similar to label propagation, we track all parts that appear in the neighborhood ($isAssigned$); however, unlike label propagation, we randomly select one of these parts instead of assigning to $v$ the part that has the maximal count among $v$'s neighbors. In practice, doing so tends to result in slightly more balanced partitions at the end of initialization.

A thread-local queue $Q_{\mathit{thread}}$ is used to maintain any new part assignment to thread-owned vertices. All threads update a MPI task-level queue which is used in ExchangeUpdates(). ExchangeUpdates() also returns a queue of updates $Q_{recv}$ for the local task's ghost vertices. We describe ExchangeUpdates() in Algorithm~\ref{alg:exchange_updates}. Algorithm~\ref{alg:pulp_init} iterates as long as tasks have updated part assignments. The number of iterations needed is on the order of the graph diameter, which can be very large for certain graph classes (e.g. road networks), leading to long execution times for this initialization stage. However, for the small-world networks that we are considering, this issue is minimal. For other graph classes, alternative strategies such as random or block assignments can be used.


\begin{algorithm}[htb]
\verysmallfont
\begin{algorithmic}[htb]
\State $Q_{recv} \gets$ ExchangeUpdates($G(V, E)$, $\mathit{parts}$, $Q_{task}$)
\State $\mathit{procid} \gets$ localTaskNum()
\State $\mathit{nprocs} \gets$ numTasksMPI()
\State $sendCounts(1 \ldots \mathit{nprocs}) \gets 0$
\ForAll{$v \in Q_{task}$} \Comment{across threads}
  \State $toSend(1 \ldots \mathit{nprocs}) \gets \textbf{false}$
  \ForAll{$\langle v, u\rangle \in E$}
    \State $task \gets$ getTask($u$)
    \If{$task \neq \mathit{procid} $ \textbf{and} $toSend(task) = \textbf{false}$}
      \State $toSend(task) = \textbf{true}$
      \State $sendCounts(task) \gets sendCounts(task) + 2$
    \EndIf
  \EndFor
\EndFor
\State {\it sendOffsets}$(1 \ldots \mathit{nprocs}) \gets$ prefixSums($sendCounts$)
\State {\it tmpOffsets} $ \gets $ {\it sendOffsets}
\ForAll{$v \in Q_{task}$} \Comment{across threads}
  \State $toSend(1 \ldots \mathit{nprocs}) \gets \textbf{false}$
  \ForAll{$\langle v, u\rangle \in E$}
    \State $task \gets$ getTask($u$)
    \If{$task \neq \mathit{procid} $ \textbf{and} $toSend(task) = \textbf{false}$}
      \State $toSend(task) = \textbf{true}$
      \State {\it sendBuffer(tmpOffsets(task))} $\gets v$
      \State {\it sendBuffer(tmpOffsets(task)+1)} $ \gets \mathit{parts}(v)$
      \State {\it tmpOffsets(task)} $ \gets $ {\it tmpOffsets(task)+2}
    \EndIf
  \EndFor
\EndFor
\State Alltoall($sendCounts$, $recvCounts$)
\State {\it recvOffsets}$(1 \ldots \mathit{nprocs}) \gets$ prefixSums($recvCounts$)
\State Alltoallv({\it sendBuffer}, $sendCounts$, {\it sendOffsets},
\State ~~~~~~~~~~~~~~~~~~~~~~~~~$Q_{recv}$, $recvCounts$, {\it recvOffsets})
\end{algorithmic}
\caption{\Xtrapulp Communication Routine:  \newline
$Q_{recv} \gets$ ExchangeUpdates($\mathit{parts}$, $Q_{task}$, $G(V, E)$)}
\label{alg:exchange_updates}
\end{algorithm}

\paragraph{\bf{ExchangeUpdates}}
This method does an Alltoallv exchange into $Q_{recv}$. Each task creates an array ($sendCounts$) for the number of items sent to other tasks and an array {\it sendOffsets}) that has start offsets for the items being sent in the send buffer.  $sendCounts$  is updated by examining all $v$ in $Q_{task}$ that have updated part assignments in the current iteration. The vertex and new part assignment is sent to any process in its neighborhood. We use the boolean array $toSend$ to avoid redundant communication. A prefix sum on $sendCounts$ yields {\it sendOffsets}.

A temporary copy of {\it sendOffsets} ({\it tmpOffsets}) is used to loop through $Q_{task}$ to fill the send buffer {\it sendBuffer}. Both loops through $Q_{task}$ can use thread-level parallelism. The updates to the buffer, offsets, and counts arrays can either be done atomically or with thread local arrays synchronized at the end. Our implementation does the latter as it shows better performance in practice. Once {\it sendBuffer} is ready, an Alltoall exchange of $sendCounts$ allows to find the number of items each task will receive ($recvCounts$). We use $recvCounts$ to create an offsets array {\it recvOffsets} for the receiving buffer $Q_{recv}$. With all six arrays initialized, an Alltoallv exchange can be completed.

\subsection{\Xtrapulp Vertex Balancing Phase}

There can be considerable imbalance after the initialization phase. The vertex balancing stage of \Xtrapulp utilizes label propagation with a weighting function $W_v$ to achieve the balance objective.  $W_v$ is roughly proportional to the target part size $\mathit{Imb}_v$ divided by the estimated current part size; its value changes as vertices are assigned to parts. We highlight the primary differences of our algorithm (Algorithm~\ref{alg:pulp_v}) from the shared memory version~\cite{pulp} here. We omit details for brevity, but the reasoning behind the calculation and updating the baseline weighting function $W_v$ was provided previously~\cite{pulp}.

\begin{algorithm}[!t]
\verysmallfont
\begin{algorithmic}[htb]
\State $\mathit{nprocs} \gets$ numTasksMPI()
\State $S_v(1 \ldots p) \gets$ numVertsPerPart($1 \ldots p$)
\State $C_v(1 \ldots p) \gets 0$
\State $iter \gets 0$
\While{$iter < I_{\text{bal}}$}
  \State $Max_v \gets $Max($S_v(1 \ldots p$), $\mathit{Imb}_v$)
  \State $\mathit{mult} \gets \mathit{nprocs}\times((X-Y)(\frac{iter_{\text{tot}}}{I_{\text{tot}}})+Y)$
  \For{$i = 1 \ldots p$}
    \State $W_v(i) \gets$ Max($\mathit{Imb}_v / (S_v(i) + \mathit{mult}\times C_v(i))- 1$, $0$)
  \EndFor
  \ForAll{$v \in V$} \Comment{across threads}
    \State $counts(1 \ldots p) \gets 0$
    \ForAll{$\langle v, u\rangle \in E$}
      \State {\it counts}$(\mathit{parts}(u)) \gets $ {\it counts}$(\mathit{parts}(u))$ + degree($u$)
    \EndFor
    \For{$i = 1 \ldots p$}
      \If {$S_v(i) + \mathit{mult}\times C_v(i) + 1 > Max_v$}
        \State $counts(i) \gets 0$
      \Else
        \State $counts(i) \gets counts(i) \times W_v(i)$
      \EndIf
    \EndFor
    \State $x \gets \mathit{parts}(v)$
    \State $w \gets$ Max($counts(1\ldots p)$)
    \If{$x \neq w$}
      \State Update($C_v(x)$,$C_v(w)$) \Comment{atomic update}
      \State Update($W_v(x)$,$W_v(w)$)
      \State $\mathit{parts}(v) \gets w$
      \State $Q_{\mathit{thread}} \gets \langle v, w\rangle$
    \EndIf
  \EndFor
  \State $Q_{task} \gets Q_{\mathit{thread}}$ \Comment{merge thread into task queue}
  \State $Q_{recv} \gets$ ExchangeUpdates($\mathit{parts}$, $Q_{task}$, $G$)
  \ForAll{$\langle v, w\rangle \in Q_{recv}$} \Comment{across threads}
    \State $\mathit{parts}(v) \gets w$
  \EndFor
  \State Allreduce($C_v$, \texttt{SUM})
  \For{$i = 1 \ldots p$}
    \State $S_v(i) \gets S_v(i) + C_v(i)$
  \EndFor
  \State $iter \gets iter + 1$
  \State $iter_{\text{tot}} \gets iter_{\text{tot}} + 1$
\EndWhile
\end{algorithmic}
\caption{\Xtrapulp Vertex Balancing Phase: \newline
$parts \gets$ \Xtrapulp-VertBalance($G(V, E), \mathit{parts}, I_{\text{bal}}, \mathit{Imb}_v$)}
\label{alg:pulp_v}
\end{algorithm}

There are a few major differences between our work and that of prior methods. We do not explicitly update the current sizes of each part $i$ ($S_v(i)$) in each iteration of the algorithm. Instead, we calculate the number of vertices gained or lost ($C_v(i)$) in each task $i$ in the current iteration. When updating the weights applied to each task $i$ ($W_v(i)$), we find an approximate size of each part based on its size at the end of the previous iteration, the number of changes during this current iteration, and a dynamic multiplier $\mathit{mult}$. The approximate size for part $i$ is calculated as: $$S_v(i) + \mathit{mult}\times C_v(i)$$

This multiplier allows fine-tuned control of imbalance when running on thousands of processors in distributed-memory. This was not an issue in previous shared-memory work. The basic idea is to use the multiplier to limit how many new vertices a single task can add to a part. This prevents all tasks from calculating a high $W_v$ value for a presently underweight part and reassigning a large number of new vertices to that part (as there is no communication before the assignment). As the iterations progress, we linearly tighten the limit on how many updates a task can do to each part, until a final iteration, where each task can provide only up to a share of $\frac{1}{\mathit{nprocs}}(\mathit{Imb}_v - S_v(i))$ additional new vertex assignments to part $i$. This prevents the imbalance constraint from being violated for any currently balanced part. The multiplier is computed as

$$\mathit{mult} \gets \mathit{nprocs}\times((X-Y)(\frac{iter_{\text{tot}}}{I_{\text{tot}}})+Y),$$

\noindent where $iter_{\text{tot}}$ is a counter of iterations performed  across each of the two outer loops in Algorithm~\ref{alg:pulpmm}, $I_{\text{tot}}$ is the maximum number of iterations allowed, and $X$ and $Y$ are input parameters. The function $\mathit{mult}$ is a linear function with $y$ intercept (iteration 0) of ($\mathit{nprocs}\times Y$) and a final value (iteration $I_{\text{tot}}$) of ($\mathit{nprocs}\times X$). We use values of $Y = 0.25$ and $X = 1.0$, which correspond to each task being allowed to add up to $4\times$ its ``share'' of updates to a task at an initial iteration and just its ``share'' at the final iteration. 

We will show experimentally that $X$ and $Y$ values close to zero results in wild imbalance swings, as the currently most underweight part can get a high number of new vertices from each task, becoming overweight. We will also show that $X$ and $Y$ values greater than about $1.5$ can have a large negative impact on overall partition quality in terms of edge cut. Additionally, it is usually desired that $X$ be larger than $Y$, as this allows a larger number of part assignment updates on initial iterations to improve overall cut while limiting updates on later iterations to allow a lower final imbalance.


\subsection{\Xtrapulp Refinement Phase}

After $I_{\text{bal}}$ iterations of the balancing phase, \Xtrapulp uses $I_{\text{ref}}$ iterations of the refinement phase (Algorithm~\ref{alg:pulp_v_ref}). The refinement phase greedily minimizes the global number of cut edges without exceeding the vertex target part size $\mathit{Imb}_v$ (if the constraint has been satisfied during the balancing phase) or without increasing the size of any part greater than the current most imbalanced part Max($S_v(1 \ldots p)$). This algorithm can be considered a variant of FM-refinement~\cite{fm} or a constrained variant of baseline label propagation. The refinement algorithm is similar to the balancing algorithm, except that the $counts$ array is not weighted. Instead, the part of vertex $v$ will be the part assigned to most of its neighbors (similar to label propagation), with the restriction that moving $v$ to that part won't increase the parts size (or estimated size with the multiplier) to larger than $Max_v$. 


\begin{algorithm}[!t]
\verysmallfont
\begin{algorithmic}[htb]
\State $\mathit{nprocs} \gets$ numTasksMPI()
\State $S_v(1 \ldots p) \gets$ numVertsPerPart($1 \ldots p$)
\State $C_v(1 \ldots p) \gets 0$
\State $iter \gets 0$ 
\While{$iter < I_{\text{ref}}$}
  \State $Max_v \gets $Max($S_v(1 \ldots p$), $\mathit{Imb}_v$)
  \State $\mathit{mult} \gets \mathit{nprocs}\times((X-Y)(\frac{iter_{\text{tot}}}{I_{\text{tot}}})+Y)$
  \ForAll{$v \in V$} \Comment{across threads}
    \State $counts(1 \ldots p) \gets 0$
    \ForAll{$\langle v, u\rangle \in E$}
      \State $counts(\mathit{parts}(u)) \gets counts(\mathit{parts}(u)) + 1$
    \EndFor
    \For{$i = 1 \ldots p$} 
      \If {$S_v(i) + \mathit{mult}\times C_v(i) + 1 > Max_v$}
        \State $counts(i) \gets 0$
      \EndIf
    \EndFor
    \State $x \gets \mathit{parts}(v)$
    \State $w \gets$ Max($counts(1\ldots p)$)
    \If{$w \neq x$}
      \State Update($C_v(x)$,$C_v(w)$) \Comment{atomic updates}
      \State $\mathit{parts}(v) \gets w$
      \State $Q_{\mathit{thread}} \gets \langle v, w\rangle$
    \EndIf
  \EndFor
  \State $Q_{task} \gets Q_{\mathit{thread}}$ \Comment{merge thread into task queue}
  \State $Q_{recv} \gets $ ExchangeUpdates($G$, $\mathit{parts}$, $Q_{task}$)
  \ForAll{$\langle v, w\rangle \in Q_{recv}$} \Comment{across threads}
    \State $\mathit{parts}(v) \gets w$ 
  \EndFor
  \State Allreduce($C_v$, \texttt{SUM})
  \For{$i = 1 \ldots p$}
    \State $S_v(i) \gets S_v(i) + C_v(i)$
  \EndFor
  \State $iter \gets iter + 1$
  \State $iter_{\text{tot}} \gets iter_{\text{tot}} + 1$
\EndWhile
\end{algorithmic}
\caption{\Xtrapulp Refinement Phase:  \newline
$\mathit{parts} \gets$ \Xtrapulp-VertRefine($G(V, E), \mathit{parts}, I_{\text{ref}}, \mathit{Imb}_v$)}
\label{alg:pulp_v_ref}
\end{algorithm}

\subsection{\Xtrapulp Edge Balancing Phase}

After $I_{\text{outer}}$ iterations of vertex balance-refinement phases, the edge balance-refinement stages begin. We don't show these algorithms for brevity, but instead will describe their differences from Algorithms~\ref{alg:pulp_v} and~\ref{alg:pulp_v_ref}. For these algorithms, we use both the target number of edges ($\mathit{Imb}_e$) and vertices ($\mathit{Imb}_v$) per task. The goal is to balance the number of edges per task while not creating vertex imbalance. The vertex weighting terms $W_v(1 \ldots p)$ are replaced by edge and cut imbalance weighting terms $W_e(1 \ldots p)$ and $W_c(1 \ldots p)$ using the current global maximum edge size per part $Max_e$ and maximum cut size per part $Max_c$. $W_e$ and $W_c$ are then used to highly weight parts that are currently underweight both in terms of the number of edges and cut edges. We weight the counts of part $i$ with the equation:
$${\it counts(i)} \gets {\it counts(i)} \times (R_e W_e(i) + R_c W_c(i))$$

$R_e$ and $R_c$ initially create bias (by first linearly increasing $R_e$ while holding $R_c$ fixed) for parts that are underweight in the number of edges. Once the edge balance constraint has been achieved, $R_e$ becomes fixed and $R_c$ correspondingly increases the bias to both minimize the maximum per-part edge cut and balance cut edges among parts. 


Using our multiplier for distributed-memory updates, we restrict the number of edges and cut edges transferred to any part per iteration; we use the same $X$ and $Y$ constants as before. However, in addition to tracking the vertex changes per part with $C_v$, edges ($C_e$) and cut edges changed per part ($C_c$) are also tracked and exchanged among tasks, as in Algorithm~\ref{alg:pulp_v}. Part sizes are updated in terms of vertices ($S_v$), edges ($S_e$), and cut edges ($S_c$), and are used to update the $W_e$ and $W_c$ weights (in addition to $R_e$ and $R_c$) as $S_v$ updated $W_v$. At the conclusion of $I_{\text{bal}}$ edge balancing iterations a refinement stage similar to Algorithm~\ref{alg:pulp_v_ref} is used. The only change in this stage is that we calculate $Max_v$ \emph{and} $Max_e$ and $Max_c$ and restrict movement of a vertex to any part that would increase the global maximum imbalance in terms of vertices, edges, and cut size.

\section{Experimental Setup}
\label{s:exp}

We evaluate \Xtrapulp performance on several small-world graphs. While \Xtrapulp is not designed for regular high-diameter graphs, we do evaluate performance on several mesh and mesh-like graphs. Table~\ref{table:graphs} lists the graphs used from the University of Florida Sparse Matrix Collection~\cite{UF, UbiCrawler, law1, law2}, the 10th DIMACS Implementation Challenge website~\cite{BaderMS0KW14}, the Stanford Network Analysis Platform (SNAP) website~\cite{SNAP, lj, orkut}, the Koblenz Network Collection~\cite{koblenz}, and Cha et al.~\cite{twitter}. Meshes used internally in our group are listed as InternalMeshX. Wherever applicable, we list statistics for the graphs. The maximum vertex degree and the graph diameter can have considerable performance impact on the label propagation and breadth first search steps that comprise our algorithm. For test instances that did not include an approximate diameter estimate, we estimate it using 10 iterative breadth first searches with a vertex randomly selected from the farthest level on the previous search. We treat all graph edges as undirected edges. 

\begin{table}[!t]
\verysmallfont
\centering
\caption{Test graphs: \# Vertices $n$, \# Edges $m$, Vertex degrees (average $d_{\textup{avg}}$ and max $d_{\textup{max}}$), and approximate diameter $\widetilde{\textup{D}}$ are listed.} 
\sisetup{ 
	table-number-alignment = center,
	table-figures-integer = 4,
	table-figures-decimal = 1,
	table-figures-exponent = 0,
	table-auto-round
}
\tabcolsep=0.075cm
\begin{tabular}{@{}
		l
		S
		S[table-figures-integer = 6, table-figures-decimal = 0]
		S
		S[table-figures-decimal = 3]
		S
@{}}
\toprule
 &  
\multicolumn{1}{c}{$n$} & \multicolumn{1}{c}{$m$} & 
  & \multicolumn{1}{c}{$d_{\textup{max}}$} & \\ 
\rb{\textbf{Graph}} &  
\multicolumn{2}{c}{$(\times 10^6)$} & 
\multicolumn{1}{c}{\rb{$d_{\textup{avg}}$}} & \multicolumn{1}{c}{$(\times 10^3)$} & 
\multicolumn{1}{c}{\rb{$\widetilde{\textup{D}}$}} \\ 

\midrule
lj              & 5.4 &  69  & 14 &  23  &    16 \\
orkut           & 3.1 & 117  & 38 &  33  &     9 \\ 
friendster      &  66 & 1806 & 53 & 5.2  &    32 \\
twitter         &  53 & 1963 & 38 & 780  &    19 \\
wikilinks       &  26 & 601  & 23 &  39  &   830 \\
dbpedia         &  67 & 258  &  4 & 7333 &     8 \\
\midrule
indochina       & 7.3 & 149  & 41 & 256  &    27 \\
arabic          &  23 & 552  & 49 & 576  &    48 \\
it              &  41 & 1151 & 29 & 1327 &    26 \\
sk              &  51 & 1949 & 38 & 8563 &   308 \\
uk-2002         & 1.8 & 298  & 16 & 195  &    30 \\
uk-2005         &  39 & 781  & 40 & 1776 &    21 \\
uk-2007         & 106 & 3302 & 31 & 975  &    25 \\
wdc12-pay       &  39 & 623  & 16 & 4933 &    13 \\
wdc12-host      &  89 & 2043 & 23 & 3391 &    19 \\
\midrule
rmat\_22        & 4.2 &  67  & 16 &  121 &     7 \\
rmat\_24        &  17 & 268  & 16 &  389 &     9 \\
rmat\_26        &  67 & 1074 & 16 &  670 &     9 \\
rmat\_28        & 268 & 4295 & 16 & 1153 &     9 \\
\midrule
InternalMesh1   & .274& 3.5 & 13 & 0.026 &   116 \\
InternalMesh2   & 2.2 &  28 & 13 & 0.026 &   232 \\
InternalMesh3   &  18 & 220 & 13 & 0.026 &   464 \\
InternalMesh4   & 140 & 1819& 13 & 0.026 &   631 \\
nlpkkt160       & 8.3 & 112 & 13 & 0.027 &   142 \\
nlpkkt200       &  16 & 216 & 13 & 0.027 &   203 \\
nlpkkt240       &  28 & 373 & 13 & 0.027 &   243 \\
\midrule
\multicolumn{5}{l}{\emph{Graphs used for Blue Waters strong scaling runs}}\\
WDC12           & 3564 & 128373 & 36 &  95097 & 5200 \\
RMAT            & 3564 & 128290 & 36 & \multicolumn{1}{c}{-} & \multicolumn{1}{c}{-} \\
RandER         & 3564 & 128290 & 36 & \multicolumn{1}{c}{-} & \multicolumn{1}{c}{-} \\
RandHD  & 3564 & 128290 & 36 & \multicolumn{1}{c}{-} & \multicolumn{1}{c}{-} \\
\midrule
\multicolumn{5}{l}{\emph{Graphs used for Blue Waters weak scaling runs}}\\
RMAT           & \multicolumn{1}{c}{$2^{25}$ to $2^{34}$} & \multicolumn{1}{c}{$2^{29}$ to $2^{40}$} & \multicolumn{1}{c}{16/32/64} &  &  \\
RandER             & \multicolumn{1}{c}{$2^{25}$ to $2^{34}$} & \multicolumn{1}{c}{$2^{29}$ to $2^{40}$} & \multicolumn{1}{c}{16/32/64} &  &  \\
RandHD             & \multicolumn{1}{c}{$2^{25}$ to $2^{34}$} & \multicolumn{1}{c}{$2^{29}$ to $2^{40}$} & \multicolumn{1}{c}{16/32/64} &  &  \\
\bottomrule
\end{tabular}
\label{table:graphs}
\end{table}

Table~\ref{table:graphs} has six sections. The first section lists four graphs that are snapshots of online social and communication networks (lj, orkut, friendster, twitter) and two hyperlink graphs (wikilinks, dbpedia). The next section includes nine web crawls of various sizes. The third section lists synthetic graphs generated using the R-MAT graph model~\cite{rmat}. The fourth section lists regular scientific computing graphs. We perform large-scale evaluations on the 2012 Web Data Commons hyperlink graph\footnote{ http://webdatacommons.org/hyperlinkgraph/}, which is created from the Common Crawl web corpus\footnote{http://commoncrawl.org}. This graph contains 3.56 billion vertices and 128 billion edges, and is the largest publicly available real-world graph known to us. For performance and scaling comparisons, we also use R-MAT (labeled RMAT) and Erd\"{o}s-R\'{e}nyi (labeled RandER) random graphs. Additionally, we generate random graphs with a high diameter (labeled RandHD) by adding edges using the following procedure: for a vertex with identifier $k$,  $0 \le k < n$, we add $d_\text{avg}$ edges connecting it to vertices chosen uniform randomly from the interval $(k-d_{\text{avg}}, k+d_{\text{avg}})$. 

We use two compute platforms for evaluations. Cluster-1 is a 16 node cluster; each node has two eight-core 2.6 GHz Intel Xeon E5-2670 (Sandy Bridge) CPUs and 64~GB main memory. We also used the NCSA Blue Waters supercomputer for large-scale runs. Blue Waters is a Cray XE6/XK7 system with \SI{22640} XE6 compute nodes and \SI{4228} XK7 compute nodes. We used only the XE6 nodes. Each node has two eight-core 2.45 GHz AMD Opteron 6276 (Interlagos) CPUs and 64~GB memory. Our experiments used up to \SI{8192} nodes of Blue Waters, which is about 36\% of the XE6 total capacity.

\section{Results}
\label{s:results}

We will demonstrate the performance of our new partitioner by assessing its scalability, partition quality, and impact on distributed graph analytics. We compare against ParMETIS version 4.0.3~\cite{parmetis} and \Pulp version 0.1~\cite{pulp}. We used the default settings of ParMETIS and \Pulp for all experiments. The build settings (C compiler, optimization flags, MPI library) for all the codes were similar on Blue Waters and Cluster-1. Unless otherwise specified, we use one MPI task per compute node for multi-node parallel runs of \Xtrapulp, and set the number of OpenMP threads to the number of shared-memory cores. 

\subsection{Performance and Scalability}

\subsubsection{Scaling on Blue Waters}

\begin{figure}[!t]
\centering
\includegraphics[width=0.35\textwidth]{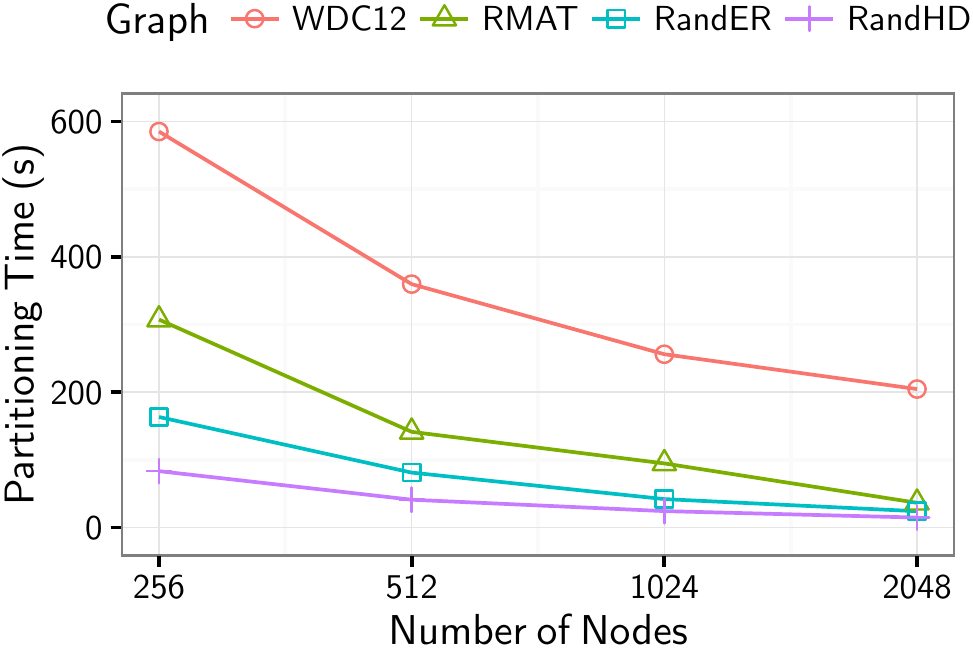}
\caption{\Xtrapulp parallel performance (\emph{strong scaling}) results on Blue Waters for computing 256 parts of various test graphs.}
\label{fig:strong_bw}
\end{figure}

We first analyze \Xtrapulp performance when running in a massively parallel setting on the Blue Waters supercomputer. Figure~\ref{fig:strong_bw} gives the execution time for partitioning the real-world Web Data Commons hyperlink graph (WDC12) and three generated graphs (RMAT, RandER, RandHD) of nearly the same size (3.56 billion vertices and 128 billion edges). We run on 
256-2048 nodes of Blue Waters (4096-32768 cores), and compute 256 parts. 

As shown in Figure~\ref{fig:strong_bw}, \Xtrapulp exhibits good \emph{strong scaling} up to 2048 nodes on all tested graphs. The speedups achieved are 2.9$\times$, 8.4$\times$, 6.8$\times$, and 5.7$\times$ for WDC12, RMAT, RandER, and RandHD graphs, respectively, when going from 256 to 2048 nodes (8$\times$ increase in parallelism). As expected, we see better speedups for the synthetic graphs due to better computational and communication load balance. The running times depend on the initial vertex ordering. The partitioning time for the RandHD network on 256 nodes is nearly $\frac{1}{7}$ the partitioning time for WDC12, even though the graphs are the same size. This is due to significantly lower inter-node communication time (which relates to the initial edge cut) in both the vertex and edge balancing steps.

\begin{figure}[!t]
\centering
\includegraphics[width=0.48\textwidth]{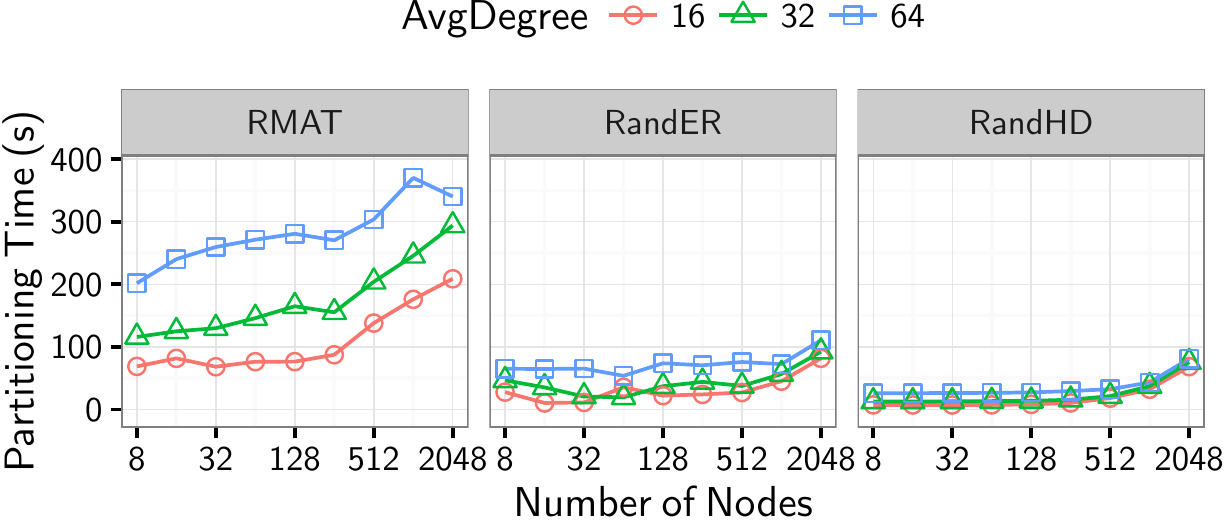}
\caption{\Xtrapulp parallel performance (\emph{weak scaling}) results on Blue Waters for various RMAT, RandER, and RandHD graphs. The number of graph vertices per node is $\approx 2^{22}$. The number of parts computed is set to number of nodes.}
\label{fig:weak_bw}
\end{figure}

Next, we perform \emph{weak scaling} experiments on Blue Waters, using 8 to 2048 compute nodes. We generate RMAT, RandER, and RandHD graphs of different sizes, and double the number of vertices as the node count doubles. The 8-node runs use graphs with $2^{25}$ vertices, whereas the 2048-node runs are for graphs with $2^{33}$ vertices. We also vary the average vertex degree, using $d_{avg} = $ 16, 32, and 64. The number of parts computed is set to the number of nodes being used for the run; thus, the computational cost changes as more parts are computed when the number of nodes increases. Figure~\ref{fig:weak_bw} shows these results. We see that partitioning time is lowest for RandHD and highest for RMAT, similar to the strong scaling results. RMAT graphs appear to be the most sensitive to average degree (or edge count) variation. For 2048-node runs, when increasing the average degree (and thereby, the number of edges) by 4$\times$ (16 to 64), the running times of RMAT, RandER, and RandHD graphs increase by 1.63$\times$, 1.35$\times$, and 1.18$\times$, respectively. Finally, we note that overall weak scaling performance is dependent on the graph structure. For the regular RandHD graphs, we see almost flat running times up to 1024 nodes, but for RMAT graphs, we observe a rise in times beyond 256 nodes. As graph size increases in RMAT graphs, so does the maximum degree, and vertices with high degrees lead to computation imbalance with the one-dimensional graph distribution used in \Xtrapulp. 

\subsubsection{Trillion Edge Runs}

We ran additional experiments on up to 8192 nodes, or 131072 cores of Blue Waters, with synthetically generated graphs with up to 17 billion vertices and 1.1 trillion edges. These tests use over a third of the available compute nodes on Blue Waters. At this scale, communication time tends to dominate the overall running time, and network traffic can have a considerable impact on total execution time. We were able to partition 17 billion ($2^{34}$) vertex, 1.1 trillion ($2^{40}$) edge RandER and RandHD graphs in 380 seconds and 357 seconds, respectively, on 8192 nodes. The largest RMAT graph we could partition on 8192 nodes had half as many edges ($2^{34}$ vertices and $2^{39}$ edges); it took 608 seconds. It should be noted that parallel partitioning is partly a ``chicken-and-egg'' problem: in order to further improve weak scaling performance for, say, RMAT graphs, we would need to statically reduce inter-node data volumes exchanged, which is dependent on the initial vertex ordering. \Xtrapulp strong and weak scaling results on Blue Waters demonstrate that there are no performance-crippling bottlenecks at scale in our implementation. 

\subsubsection{Scaling on Cluster-1}

We extensively test \Xtrapulp at a smaller scale (16 nodes of Cluster-1), for direct performance comparisons to ParMETIS and \Pulp. \Xtrapulp exploits hybrid MPI and thread-level parallelism, and we use a single MPI task per compute node. For MPI-only ParMETIS, we run 16, 8, 4, and 1 tasks per node and report the best time in order to provide a conservative comparison. OpenMP-only \Pulp results are with full threading on a single node. Note that \Xtrapulp is explicitly designed for much larger-scale processing, so we perform this small-scale analysis here only to give relative performance comparisons to the current state-of-the-art.

\begin{table}[!t]
	\supersmallfont
	\centering
	\caption{\Xtrapulp, \Pulp, and ParMETIS parallel performance results on Cluster-1 for computing 16 parts of various test graphs. \Xtrapulp and \Pulp results include 16-way multithreaded parallelism, and ParMETIS results are the best ones obtained with 16- to 256-way MPI task concurrency. $\dagger$/$\ddagger$ symbols indicates relative speedup with respect to 2/4-node \Xtrapulp runs.}
	\sisetup{ 
		table-number-alignment = center,
		table-figures-integer = 4,
		table-figures-decimal = 2,
		table-figures-exponent = 0,
		table-auto-round
	}
	\tabcolsep=0.04cm
	\begin{tabular}{@{}lSSS
			S[table-figures-integer = 3, table-figures-decimal = 1]
			S[table-figures-integer = 3, table-figures-decimal = 1]
			@{}}
		\toprule
		& \multicolumn{3}{c}{\textbf{Partitioning Time (s)}} & \multicolumn{2}{c}{\textbf{\Xtrapulp Speedup}} \\
		\cmidrule(lr){2-4} \cmidrule(lr){5-6}
		\textbf{Graph} & \multicolumn{1}{c}{\Xtrapulp} & 
		\multicolumn{1}{c}{\Pulp}& \multicolumn{1}{c}{ParMETIS}  &
		& \multicolumn{1}{c}{Rel. to} \\ 
		& \multicolumn{1}{c}{(16 nodes)} & 
		\multicolumn{1}{c}{(1 node)}& \multicolumn{1}{c}{(16 nodes)}  &
		\multicolumn{1}{c}{\rb{vs \Pulp}} & \multicolumn{1}{c}{1 node} \\ 
		
		\midrule
		lj            &   \bfseries 4.9 & 10 & 59 &      2.04$\times$ & 8.9$\times$ \\
		orkut         &   \bfseries 4.8 & 18 & 110 &       3.8$\times$ & 8.6$\times$\\
		friendster    &   \bfseries 232 &  1672 & &    \bfseries 7.2$\times$ & 11$\times$\\ 
		twitter       &   \bfseries 1647  & 3611  & &  2.2$\times$ & 2.3$^{\dagger}\times$\\
		wikilinks     &  \bfseries 137 &    467 & &    3.4$\times$ & 5.9$\times$\\
		dbpedia       &  \bfseries 35 &  70 &  &       1.999$\times$ & \bfseries 14$\times$\\
		\midrule
		indochina     &  \bfseries 4.4 &  8.1  &    130 &    1.84$\times$ & 11.252$\times$\\
		arabic        &  \bfseries 12 &   16 & 754 &     1.3$\times$ & 8.173$\times$\\
		it            & \bfseries 22 &  32 &  &        1.4$\times$ & 9.389$\times$\\  
		sk            & \bfseries 33 & 67 &    &       2.1$\times$ & 9.094$\times$\\ 
		uk-2002       &  \bfseries 5.1 & 9.2 & 85 &      1.8$\times$ & \bfseries 12.771$\times$\\
		uk-2005       &  \bfseries 18 & 34 &  &        1.9$\times$ & 9.76$\times$\\
		uk-2007       &  \bfseries 49 & 71 &  &        1.4$\times$ & 3.926$^{\dagger}\times$\\
		wdc12-pay           &   \bfseries 241 & 1062 & &     4.4$\times$ & 6.17$\times$\\
		wdc12-host          &   \bfseries 422 &  2443 & &   \bfseries 5.7$\times$ & 8.583$\times$\\
		\midrule
		rmat\_22      &  \bfseries 6.7 &   14 & 126 &     2.1$\times$ & 4.91$\times$\\
		rmat\_24      &   \bfseries 30 & 147 & 923 &      4.9$\times$ & 10.53$\times$\\
		rmat\_26      &  \bfseries 183 & 1022 &  &     5.6$\times$ & \bfseries 12.528$\times$\\
		rmat\_28      &   \bfseries 981 &  5454 &  &  \bfseries 5.6$\times$ & 3.23$^{\ddagger}\times$\\
		\midrule
		InternalMesh1 & \bfseries 0.09 & 0.14 & 0.63 &    1.556$\times$ & 20.02$\times$\\ 
		InternalMesh2 & \bfseries 0.45 &  0.89 & 0.71 &   1.97$\times$ & 23.43$\times$\\ 
		InternalMesh3 & 2.9 & 6.8 & \bfseries 1.2 &     2.34$\times$ & \bfseries 27.86$\times$\\ 
		InternalMesh4 &  24 &   46 & \bfseries 4.6 &    1.92$\times$ & 26.74$\times$\\ 
		nlpkkt160     &  1.6 & 3.8 & \bfseries 1.5 &    2.375$\times$ & 11.52$\times$\\ 
		nlpkkt200     &   2.6 &  6.4 & \bfseries 2.2 &  \bfseries 2.46$\times$ & 13.635$\times$\\ 
		nlpkkt240     &  4.6 &  11 & \bfseries 3.6 &    2.39$\times$ & 13.46$\times$\\ 
		\bottomrule
	\end{tabular}
	\label{table:best_times}
\end{table}

We present 16-node performance results in Table~\ref{table:best_times}. Empty cells in the table indicate cases where ParMETIS failed to run to completion, and this was mostly due to out-of-memory and related errors on some MPI task. We indicate in bold font the best timing results for each graph, and the best \Xtrapulp absolute speedup (with respect to single-node \Pulp) and relative speedup results in each of the four graph classes. The single-node shared-memory \Pulp is consistently faster than distributed-memory parallel ParMETIS for the first three classes of graphs. This is expected, given that the label propagation achieves good shared-memory strong scaling and has been shown to work well as a community detection and partitioning approach for small-world graphs. For the fourth class of regular, high-diameter graphs, ParMETIS outperforms \Pulp and \Xtrapulp; ParMETIS is optimized to partition these types of graphs.

For all of the small-world graphs, 16-node \Xtrapulp running times are better than single-node \Pulp running times. \Xtrapulp and \Pulp have several key algorithmic differences, and single-node \Pulp is faster than single-node \Xtrapulp. We omit a direct comparison between single node performance, but these values can be inferred through the last two columns in the table. We created \Xtrapulp in order to scale to multi-node settings, and we see that the 16-node speedup (with respect to single-node \Xtrapulp) is quite good, being 14$\times$ for dbpedia 12.8$\times$ for uk-2002. The speedup relative to \Pulp is also quite good, considering the difficulties and overheads in reformulating an asynchronous shared-memory algorithm into a synchronous distributed-memory implementation. E.g., in the current (June 2016) version of the graph500.org benchmark, the per-core performance ratio between the fastest shared-memory implementation and fastest distributed-memory implementation is approximately 6.5$\times$; our ratios are of a similar order, being between 11$\times$ for it and uk-2007 and only 2.2$\times$ for friendster, despite our implementation not being as finely optimized as the Graph500 benchmark code. 

For large graphs such as wdc12-host, we achieve a significant reduction in running time by exploiting multiple compute nodes. We also note that the superlinear relative speedups for InternalMesh graphs are due to the fact that the initial partitioning of these graphs is actually quite good, thereby leading to a low communication-to-local computation ratio.




\begin{figure}[!t]
\centering
\includegraphics[width=0.35\textwidth]{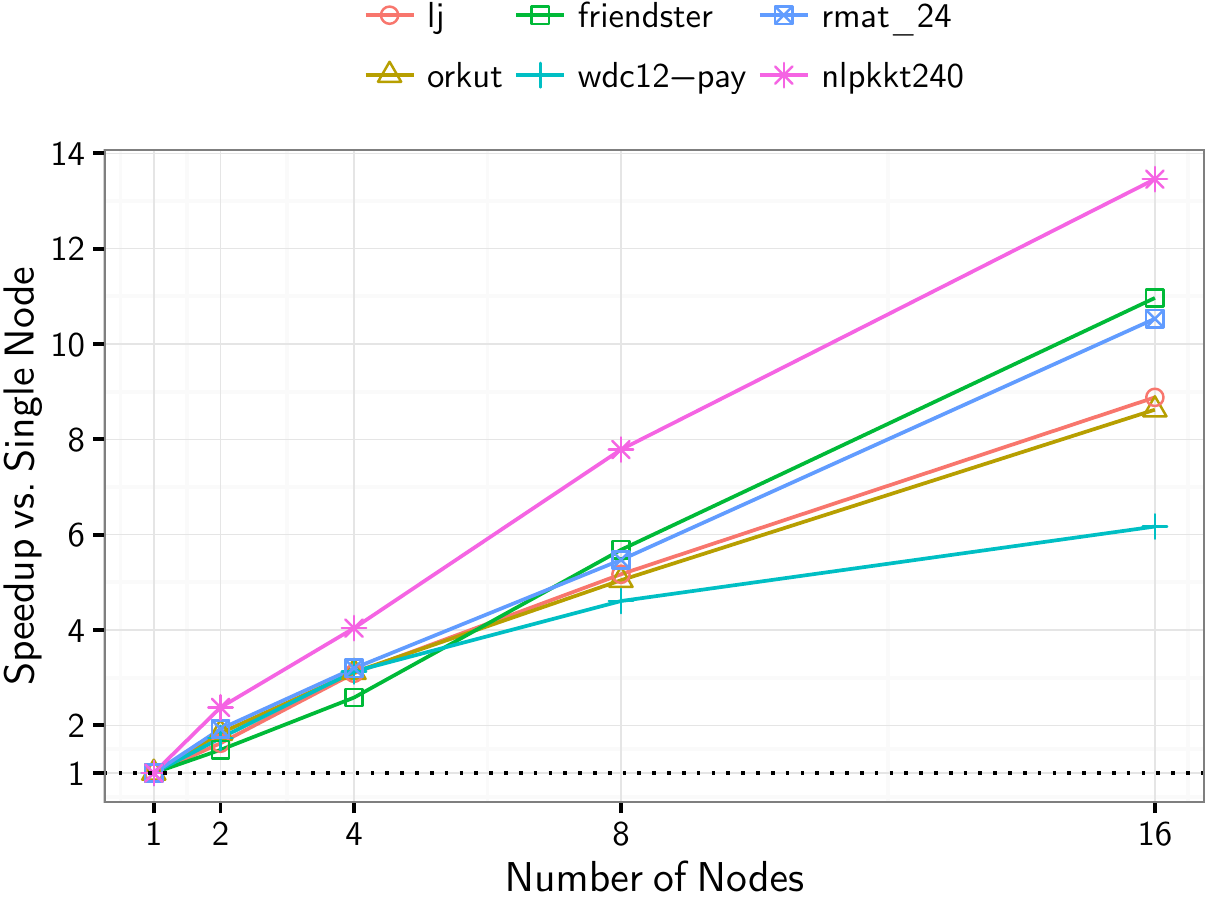}
\caption{
	\Xtrapulp relative speedup results on Cluster-1 for computing 16 parts of various graphs.}
\label{fig:strong_compton}
\end{figure}

Figure~\ref{fig:strong_compton} shows \Xtrapulp strong scaling for six representative graphs. Note that graph sizes vary significantly, ranging from the 69 million-edge lj graph to the 1.8 billion-edge friendster graph. We observe a range of relative speedups, attributable to the graph structure.  There appear to be no intrinsic scaling bottlenecks even at this smaller 16-node scale. 



\begin{figure*}[!htb]
\centering
\includegraphics[width=0.98\textwidth]{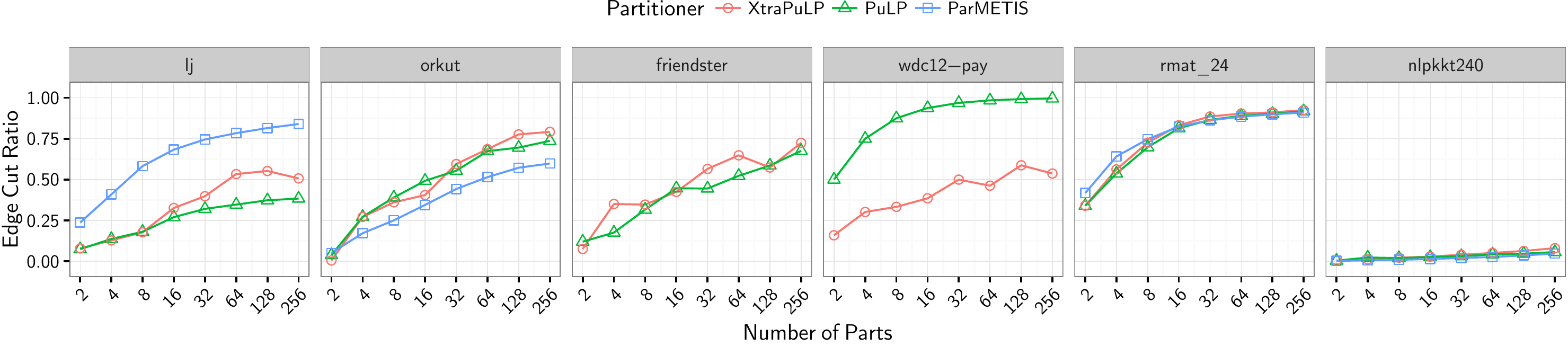}\\
\includegraphics[width=0.98\textwidth]{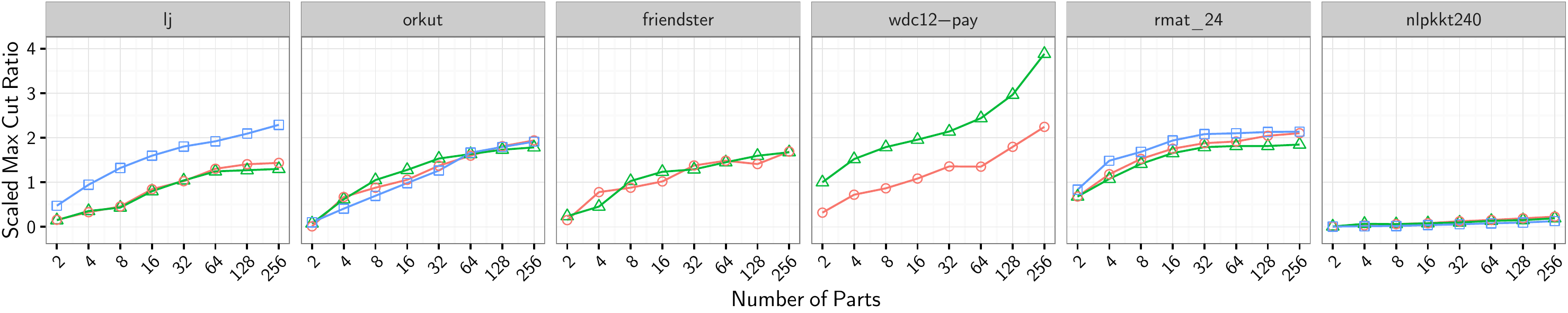}
\caption{Partition quality comparison when varying the number of parts computed. For both Edge Cut ratio and Scaled Max Cut ratio, lower values are better.}
\label{fig:quality_compton}
\end{figure*}

\subsection{Partitioning Quality}

We next evaluate \Xtrapulp partitioning quality by comparing results to \Pulp and ParMETIS. We use the two architecture-independent metrics for quality comparisons: Edge cut ratio (number of edges cut divided by the number of edges) and Scaled max edge cut ratio (maximum over all parts of the ratio of the number of edges cut to the average number of edges per part). For both metrics, lower values are preferred. These two metrics correspond to the objectives that the three partitioning methods optimize. In Figure~\ref{fig:quality_compton}, we report these metrics, varying the number of parts from 2 to 256 as quality results can vary with number of parts. We join the individual data points to indicate trends. We use the same six representative graphs that were used for strong scaling experiments on Cluster-1. Again note they we perform analysis at this scale only for relative comparison. At the scale for which \Xtrapulp is designed, the only competing methods are random and block partitioning; random partitioning produces an edge cut ratio that scales  approximately as $\frac{p-1}{p}$, where $p$ is the number of parts, while the quality of block partitioning is highly variable and dependent on how the graph is stored.


Our first observation is that both quality metrics -- edge cut ratio and scaled max cut ratio -- can vary dramatically based on the graph structure. The quality results for nlpkkt240 are in stark contrast to the rest of the graphs. On increasing the number of parts, the two metrics increase only slightly for nlpkkt240, but at a much faster rate for the rest of the graphs. The edge cut ratio quickly approaches 1.0 with increasing part count when partitioning rmat\_24. An edge cut ratio value close to 1.0 means that nearly every edge is a cut edge. For graphs with intrinsically high edge cut ratios (such as the graphs in the first class, online social networks and communication networks), any quality gains must be assessed taking partitioning running times into consideration. Ideally, the partitioning method should finish quickly for cases where quality metrics cannot be substantially improved.

Comparing \Pulp and \Xtrapulp results, we observe that the metrics are relatively close, despite the asynchronous intra-task updates in \Xtrapulp. We observe much better performance for \Xtrapulp on the wdc-pay graph, likely due to the novel initialization strategy. We also observe that \Xtrapulp trends are not as ``smooth'' as \Pulp trends for some graphs (e.g., lj, friendster), which may be due to the distributed-memory parallelization of label propagation or the iteration counts for balancing and refinement phases. ParMETIS fails to run for two of the six graphs. \Xtrapulp outperforms ParMETIS on lj, whereas ParMetis does slightly better on orkut. 


To numerically quantify quality gains/losses, we compute ``performance ratios'' for all partitioners over all tested graphs in Table~\ref{table:graphs}. Here, the performance ratio is defined as the geometric mean over all tests, of each partitioner's edge cut or max per-part cut, divided by the best edge cut or max per-part cut for that test. A lower value is better, with a ratio of exactly 1.0 indicating that the partitioner produced the best quality for every single test. We calculate performance ratios for edge cut to be 1.18, 1.33, and 1.37 and max per-part cut to be 1.19, 1.40, and 1.41 for ParMETIS, \Pulp, and \Xtrapulp, respectively. When we consider only the irregular graphs for which ParMETIS completes, the values are much closer, with edge cut ratios of 1.36, 1.36, and 1.46 and max per-part cut ratios of 1.39, 1.43, and 1.49 for ParMETIS, \Pulp, and \Xtrapulp, respectively. We thus claim that partitioning quality is not compromised for small-world graphs when using \Xtrapulp. \Xtrapulp also provides users the ability to partition large graphs that do not fit on a single node, and achieves good strong and weak scaling.

\begin{figure}[htb]
\centering
\includegraphics[width=0.15\textwidth]{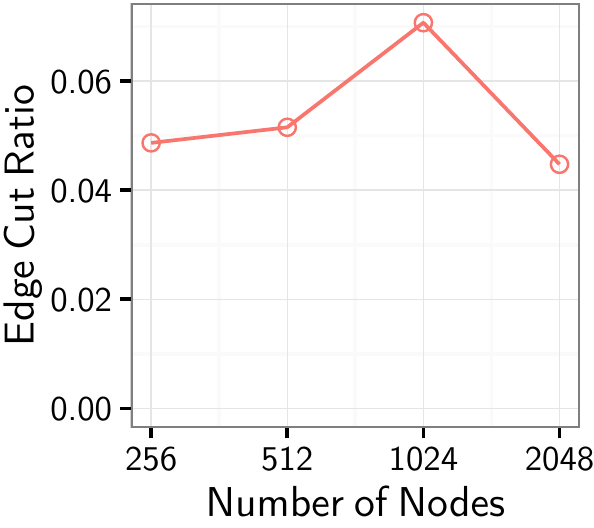}
\includegraphics[width=0.15\textwidth]{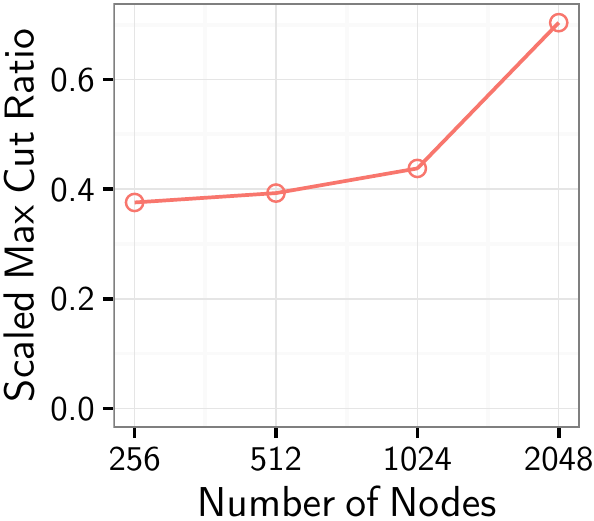}
\includegraphics[width=0.15\textwidth]{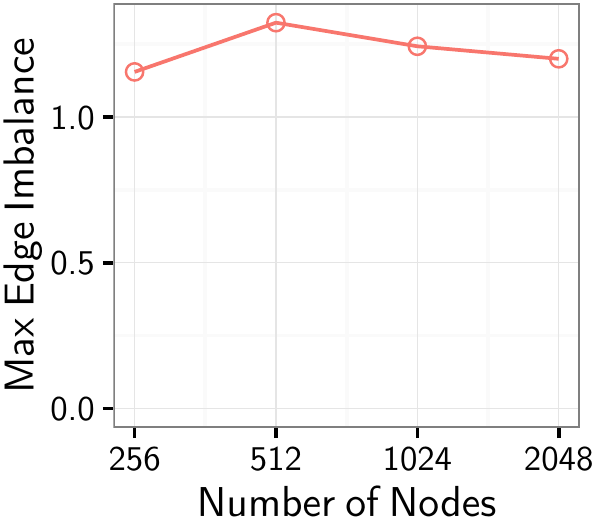}
\caption{Partitioning quality results for computing 256 parts of the WDC12 graph on Blue Waters.}
\label{fig:quality_bw}
\end{figure}

Our final quality experiment on Blue Waters measures how partition quality varies with large-scale parallelism. In Figure~\ref{fig:quality_bw}, we plot how the edge cut ratio, scaled max per-part cut ratio, and partition edge count imbalance vary with large MPI task counts, when partitioning WDC12 into 256 parts. The edge cut ratio is between 0.04 and 0.07, which is considerably lower than the values of 0.16 for vertex block partitioning and almost 1.0 for random partitioning. Note that the relatively low edge cut here for block partitioning is a result of the crawling method, but it comes at a high cost: the edge imbalance ratio is 1.85. The resulting partitions from \Xtrapulp are well-balanced. The increase in max cut ratio is possibly due to the fact that, as task count increases, the number of updates allowed per task and per iteration decreases due to the multiplier $\mathit{mult}$, introduced in the \Xtrapulp algorithms section. In future work, we will experiment further with the ($X,Y$) parameters, the multiplier, and alternate communication schemes in order to ensure more consistent partitions when scaling the number of tasks.

\subsection{Additional Comparisons}

\begin{figure}[!htb]
\centering
\includegraphics[width=0.48\textwidth]{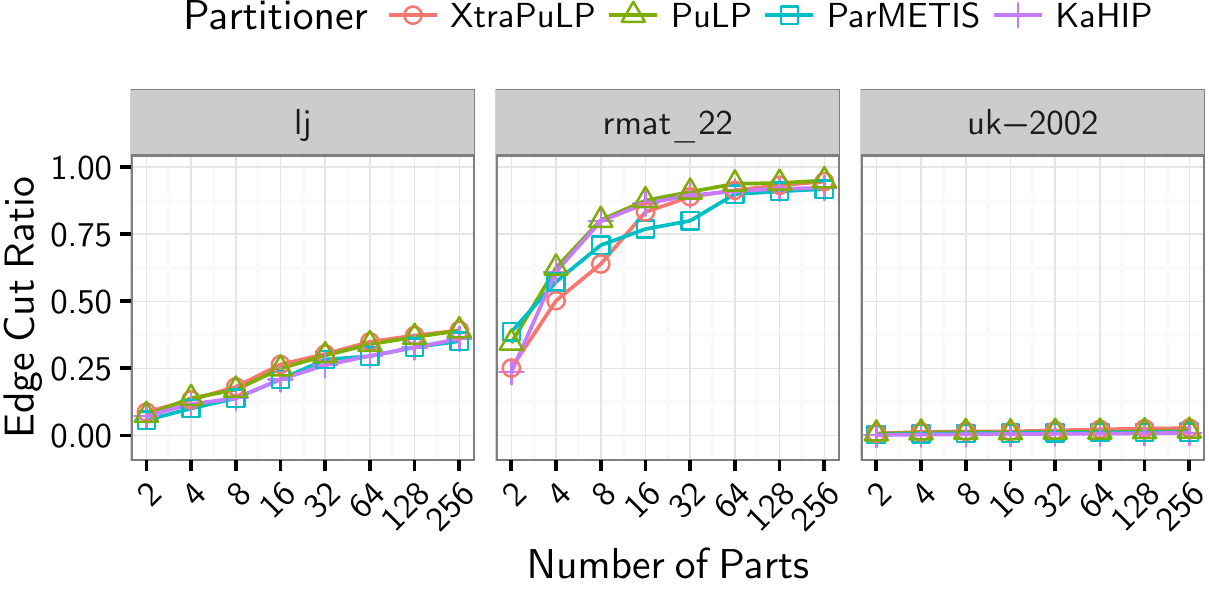}\\
\includegraphics[width=0.48\textwidth]{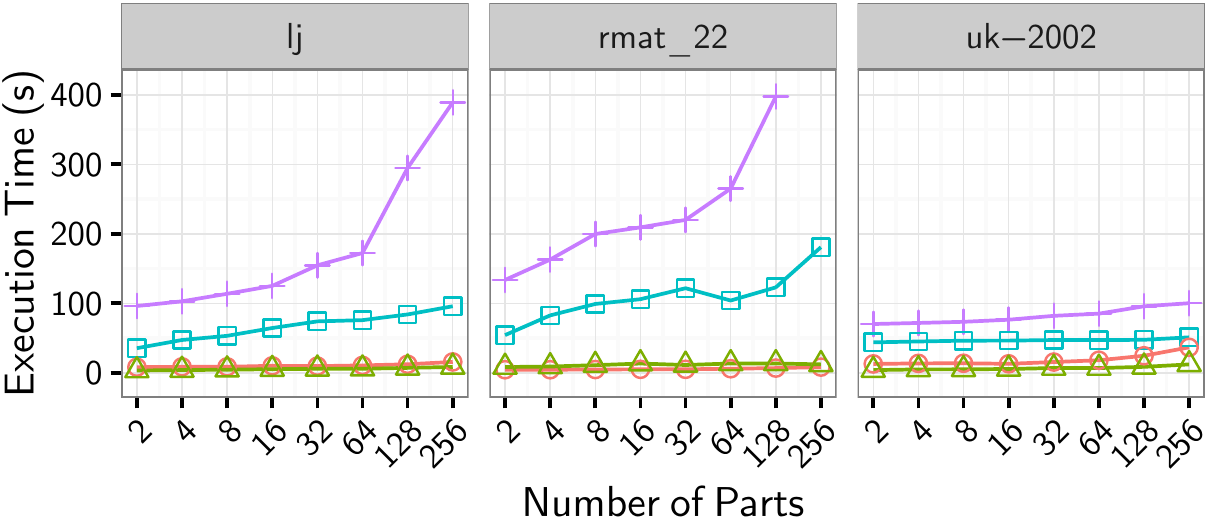}
\caption{Partitioning quality (top) and execution time (bottom) for multiple partitioners solving the single objective single constraint partitioning problem.}
\label{fig:single_obj}
\end{figure}

Here we provide additional comparisons to the recent state-of-the-art partitioner of Meyerhenke et al.~\cite{kahipPar}, which uses size-constrained label propagation during the graph coarsening phase. This partitioner solves the single-constraint and single-objective graph partitioning problem, optimizing for edge cut and balancing vertices per part. Therefore, we modify our \Xtrapulp code by eliminating the edge balancing and max per-part cut phase to provide a direct comparison. We also run shared-memory \Pulp and ParMETIS. All codes are run using 16-way parallelism with a 3\% load imbalance constraint, and we compute 2-256 parts of the lj social network, scale 22 R-MAT graph, and uk-2002 web crawl. In Figure~\ref{fig:single_obj}, we compare edge cut (top) and execution time (bottom).

Overall, we observe \Xtrapulp to be within a small fraction of the Meyerhenke et al. and ParMETIS codes in terms of part quality, while running only slightly slower than shared-memory \Pulp. This is despite \Xtrapulp being designed and optimized for the multi-objective multi-constraint problem. Performance ratios for cut quality on this limited test set are 1.05 for Meyerhenke et al., 1.23 for ParMETIS, 1.51 for \Pulp, and 1.61 for \Xtrapulp. Performance ratios for execution time were 1.27 for \Pulp, 1.73 for \Xtrapulp, 11.81 for ParMETIS, and 26.5 for Meyerhenke et al. These results demonstrate the efficiency tradeoff between quality and time to solution, the choice of which to optimize for being application-dependent.

However, we emphasize again that we provide these results only to establish a relative baseline for comparison of the performance of \Xtrapulp, as the engineering decisions driving its design were made to enable scalability to partition graphs several orders-of-magnitude larger than the graphs presented here.

\subsection{Multiplier Parameters}

\begin{figure}[!htb]
\centering
\includegraphics[width=0.23\textwidth]{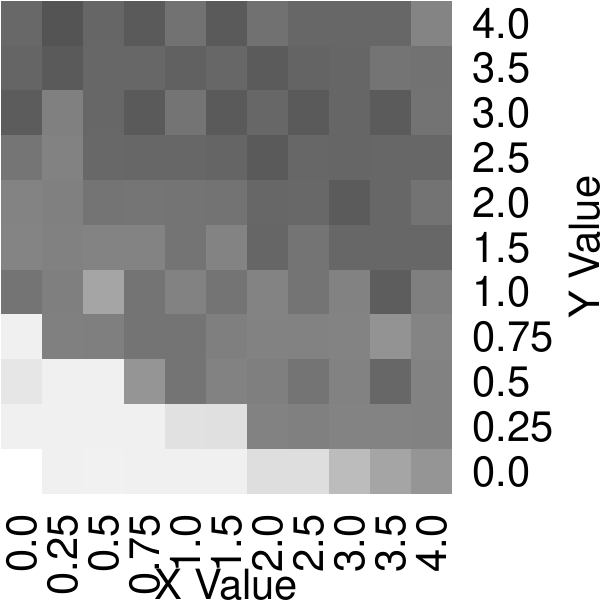}
\includegraphics[width=0.23\textwidth]{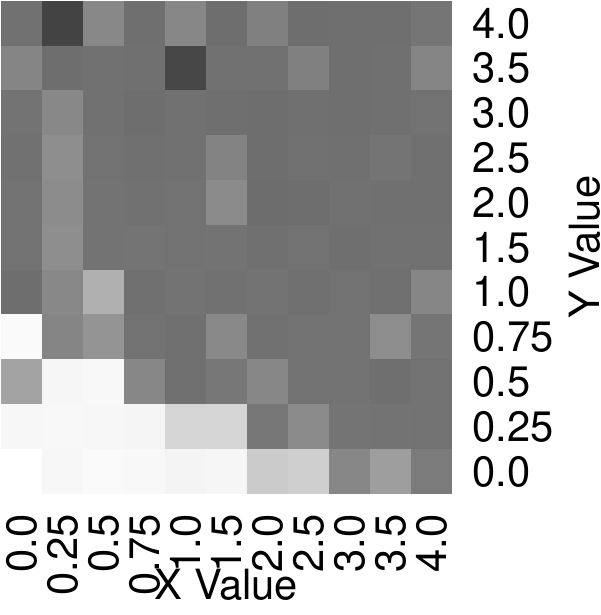}\\
\includegraphics[width=0.23\textwidth]{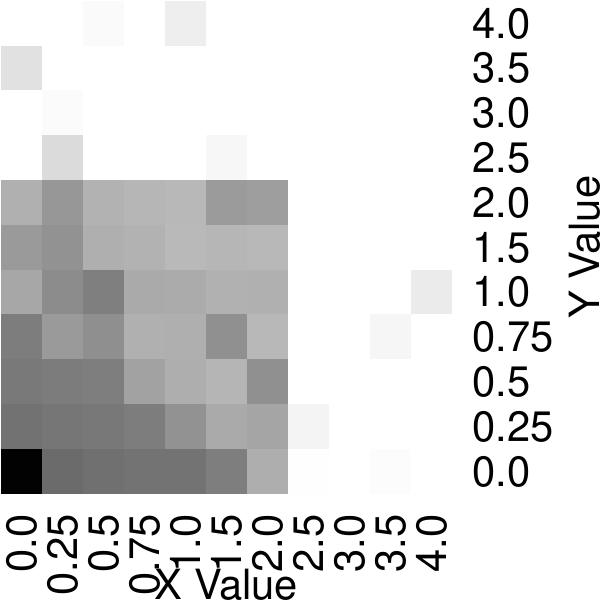}
\includegraphics[width=0.23\textwidth]{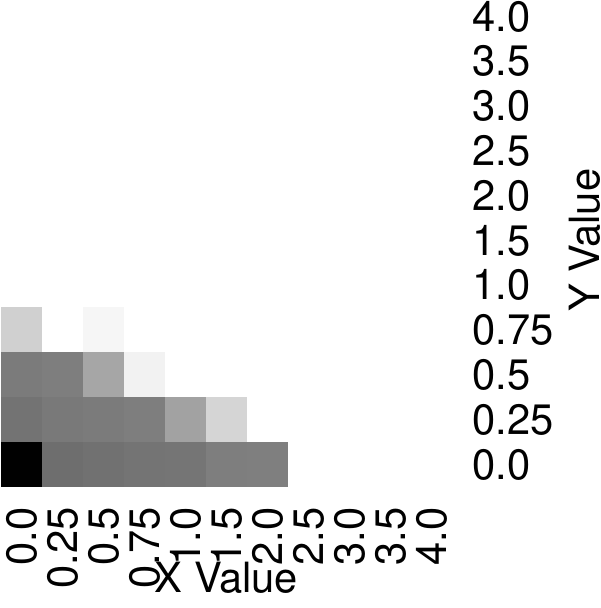}
\caption{Clockwise from upper left: Edge Cut versus $X, Y$; Max Cut versus $X, Y$; Vertex Balance versus $X, Y$; Edge Balance versus $X, Y$.}
\label{fig:quality_xy}
\end{figure}

We also analyze the effect that varying the $X$ and $Y$ parameters have on the final partition quality. Using lj, uk-2002, rmat\_22, and nlpkkt160 as representative examples for each graph class, we computed from 2-128 parts each on 2-16 compute nodes of \emph{Compton} (all powers of 2 in between). We plot heatmaps of the average results in Figure~\ref{fig:quality_xy}, with white indicated higher quality or better balance and black indicating poorer quality or balance. For the two quality plots, we omit results that exceed the 10\% balance constraint by at least 5\% or more. For the balance plots, solid white indicates that all tests conducted for that $X, Y$ achieved the balance constraints.

The top two plots give the edge cut versus $X, Y$ (left) and the max per-part cut versus $X, Y$ (right). From these plots we observe two trends. Most obviously, a lower $X$ and $Y$ indicates a higher quality cut. This is because lower values for these parameters allow the highest number of part reassignments and therefore the greatest overall refinement. Second, we notice that a higher $X$ value relative to $Y$ will, on average, also result in a better cut. This is due to how a higher initial limit on part reassignments ($Y$) and a lower final limit ($X$) can greatly refine the initial parts while limiting the potential imbalance possible on the final iterations.

The bottom two plots show overall average edge balance (left) and vertex balance (right). In general, the level of balance achieved is opposite the quality of cut. The optimal $X, Y$ pair of values should therefore be selected along the \emph{threshold}, where high quality and balance are concurrently achieved. We selected our test values of $X = 1.0$ and $Y = 0.25$ empirically, as they gave us the overall best quality in terms of cut and balance on our test suite.

\subsection{Applications}

\begin{figure}[!t]
\centering
\includegraphics[width=0.30\textwidth]{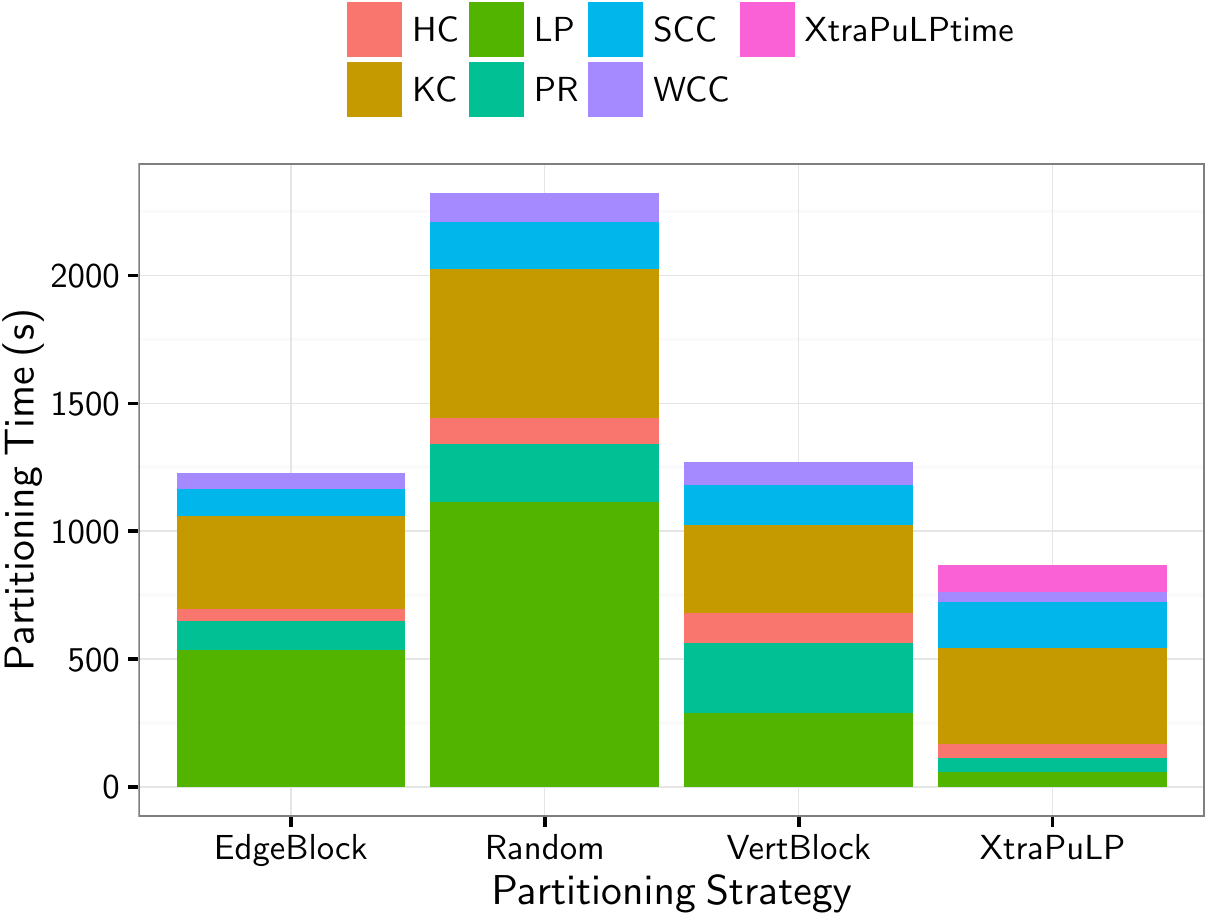}
\caption{The parallel performance results of various parallel graph analytics (HC, KC, LP, PR, SCC, WCC) on 256 nodes of Blue Waters, executed on the WDC12 graph with different graph partitioning strategies.}
\label{fig:analytics}
\end{figure}

We next demonstrate that \Xtrapulp can significantly improve performance of real-world analytics. Consider analytics on the 128 billion edge WDC12. Without a partitioner that can process graphs of this size, the common approaches to running analytics are to use simple balanced vertex and edge assignment strategies that do not optimize for edge cut. In Figure~\ref{fig:analytics}, we give the execution times of six analytics on WDC12 with four partitioning strategies. EdgeBlock partitioning stores a contiguous set of vertices and all their adjacencies in each node such that each node has approximately the same number of edges. VertexBlock partitioning stores roughly the same number of vertices and all their adjacencies in each node. Random partitioning assigns vertices to nodes randomly. \Xtrapulp assigns vertices based on the computed partition. The six analytics considered (algorithms presented in~\cite{hpcgraph}) are Harmonic Centrality (HC) computation of 100 vertices, approximate K-core decomposition (KC), Label Propagation-based community detection (LP), PageRank (PR), extraction of the largest strongly connected component (SCC), and weakly connected components decomposition (WCC). For \Xtrapulp, we \emph{include} the partitioning time in comparisons. For \Xtrapulp, we exploit prior knowledge~\cite{hpcgraph} and run the balancing stage of \Xtrapulp after first initializing with vertex block partitioning.

\begin{table*}[htbp]
	\centering
	\caption{The performance results of parallel sparse matrix vector multiplication (\textup{SpMV}) on 1 node (16 MPI tasks) to 16 nodes (256 MPI tasks) of Cluster-1, with different graph partitioning strategies. PM: \textup{ParMETIS}. We report time for 100 \textup{SpMV}s.}
	\sisetup{ 
		table-number-alignment = center,
		table-figures-integer = 4,
		table-figures-decimal = 2,
		table-figures-exponent = 0,
		table-auto-round
	}
	\tabcolsep=0.1cm
	\begin{tabular}{@{}lSSSSSSSSSS@{}}
		\toprule
		&  & \multicolumn{8}{c}{\textbf{Execution time for 100 SpMVs} (s)} & \multicolumn{1}{c}{\textbf{Speedup}} \\
		\cmidrule{3-10}
		& \multicolumn{1}{c}{MPI} & \multicolumn{4}{c}{\textbf{1D partitioning}} & \multicolumn{4}{c}{\textbf{2D partitioning}} & \multicolumn{1}{c}{2D \Xtrapulp} \\
		\multicolumn{1}{l}{\rb{\textbf{Graph}}} & \multicolumn{1}{c}{tasks}  & \multicolumn{1}{c}{Block} & \multicolumn{1}{c}{Rand} & \multicolumn{1}{c}{PM} & \multicolumn{1}{c}{\Xtrapulp} & \multicolumn{1}{c}{Block} & \multicolumn{1}{c}{Rand} & \multicolumn{1}{c}{PM} & \multicolumn{1}{c}{\Xtrapulp} & \multicolumn{1}{c}{over 1D Rand}\\
		\midrule
		lj &  16 & 19.87 & 12.24 & 10.90 & 9.81 & 15.74 & 10.09 & 9.26 & \bfseries 8.32 & 1.4711$\times$\\
		& 128 &  6.31 &  4.29 &  3.41 & 3.05 &  4.04 &  1.92 & 2.07 & \bfseries 1.85 & 2.3189$\times$\\
		& 256 &  4.73 &  2.96 &  2.65 & 2.41 &  2.32 &  1.26 & 1.28 & \bfseries 1.12 & 2.6426$\times$\\
		orkut &  16 & 28.80 & 26.62 & 23.31 & 23.82 & 19.88 & 17.29 & 17.86 & \bfseries 16.51 & 1.6123$\times$\\
		& 128 & 10.05 & 10.43 &  7.77 &  8.35 &  4.20 & \bfseries 3.07 & 3.56 &  3.42 & 3.0497$\times$\\
		& 256 & 6.55  &  6.27 &  5.63 &  5.25 &  2.25 & 1.94 & \bfseries 1.82  & 1.91  & 3.2827$\times$\\
		friendster & 128 &  240.07 & 137.21  &  & 141.72 &  161.45 & \bfseries 81.11 & & 83.33 & 1.65$\times$\\
		& 256 & 154.43  &  79.67 &  &  89.98 &  79.56 &  46.85 &  & \bfseries 46.69 & 1.706$\times$\\
		wdc12-pay    &  16 & 614.82 & 175.25 &  & 155.79 & 395.94 & 85.79 &  & \bfseries 76.99 & 2.28$\times$\\
		& 128 & 224.84 &  40.91 &  &  41.35 & 96.94  & 25.79 &  & \bfseries 24.27 & 1.6856$\times$\\
		& 256 &  153.51 & 43.14  &  &  27.61 & 59.18  &  14.65 &  & \bfseries 14.51 & 2.97$\times$\\
		rmat\_24 &  16 & 138.76 & 144.51 & 107.19 & 99.87 & 107.28 & 108.22 & 87.83 & \bfseries 83.07 & 1.74$\times$ \\
		& 128 &  44.20 &  50.49 & 27.69 & 27.28 &  25.57 &  25.31 & 15.93 & \bfseries 15.52 & 3.25$\times$ \\
		& 256 &  32.22 &  37.37 & 19.63 & 19.76 &  14.67 &  15.93 & \bfseries 9.97 & 10.02 & 3.73$\times$ \\
		nlpkkt240 &  16 & 30.96 & 26.67 & 19.86 & 17.55 & \bfseries 16.08 & 40.96 & 16.96 & 18.03 & 1.48$\times$ \\
		& 128 & 2.59  & 26.73  & \bfseries 2.14 & 2.57  &  2.71 & 14.55 & 2.24 & 2.56 & 10.44$\times$ \\
		& 256 & 1.62  & 22.61  & \bfseries 1.04 &  1.37 & 1.81  & 10.97 & 1.12 & 1.56 & 14.5$\times$\\
		\bottomrule
	\end{tabular}
	\label{tab:matvec_time}
\end{table*}

Using balanced \Xtrapulp partitions reduces end-to-end execution time by 30\%, from 1229 seconds with an edge block partitioning to 867 seconds with \Xtrapulp. We see a substantial reduction in analytics where inter-node communication time is directly proportional to total edge cut, such as PR and LP, even when including the \Xtrapulp partitioning time. We are unaware of other partitioning methods that are capable of processing graphs that are structurally similar to WDC12.

In Table~\ref{tab:matvec_time}, we present results that show partitioning impact on sparse matrix vector multiplication (SpMV). We use the Epetra package of the Trilinos scientific computing library~\cite{trilinos} to perform 100 SpMV operations, using matrices constructed from the six select test graphs previously shown. Parallel SpMV is a key computation in eigensolvers and iterative methods for liner systems. We use several partitioning strategies, including one dimensional vertex block (1D-Block), random (1D-Rand), ParMETIS (1D-PM), and \Xtrapulp (1D-\Xtrapulp). We also utilize 2D distributions with vertex block (2D-Block) and random partitions (2D-Rand). Additionally, using a strategy for mapping 1D partitions into 2D distributions~\cite{2D}, we run with 2D distributions produced from our 1D ParMETIS (2D-PM) and \Xtrapulp (2D-\Xtrapulp) partitions. We run these tests on 1, 8 and 16 nodes of Cluster-1 with 16, 128, and 256 MPI ranks, respectively. We observe that using \Xtrapulp partitioning can often accelerate the SpMV operation time in these tests. We observe a 2.77$\times$ (geometric mean) reduction in execution time when using 2D \Xtrapulp-based distributions instead of 1D-Rand for 256-way parallel code on the five irregular graphs. Regular meshes such as nlpkkt240 do not directly benefit from a 2D distribution, and hence 1D-Rand partitioning fares poorly.

\section{Conclusion}
\label{s:conc}

We introduced \Xtrapulp, our distributed-memory partitioner that can scale to graphs several orders-of-magnitude larger than prior work. This work significantly extended our prior shared-memory-only partitioner, \Pulp. We show comparable partition quality to prior methods at the small scale, and, at the large scale, we significantly improve upon the existing competing methods, block and random partitioning. We also demonstrate faster execution times and comparable parallel efficiency relative to the state-of-the-art. Using partitions computed by \Xtrapulp, we also improve performance on highly tuned matrix-vector multiplication kernels and several graph analytics running on the current largest publicly available web crawl.


\section*{Acknowledgment}
This research is part of the Blue Waters sustained-petascale computing project, which is supported by the National Science Foundation (awards OCI-0725070, ACI-1238993, and ACI-1444747) and the state of Illinois. Blue Waters is a joint effort of the University of Illinois at Urbana-Champaign and its National Center for Supercomputing Applications. This work is also supported by NSF grants ACI-1253881, CCF-1439057, and the DOE Office of Science through the FASTMath SciDAC Institute. Sandia National Laboratories is a multi-program laboratory managed and operated by Sandia Corporation, a wholly owned subsidiary of Lockheed Martin Corporation, for the U.S. Department of Energy's National Nuclear Security Administration under contract DE-AC04-94AL85000. We also thank Henning Meyerhenke, Peter Sanders, and Christian Schulz for providing the source code for their partitioner.
\vspace{-5pt}
\bibliographystyle{IEEEtranS}
\bibliography{sc16}

\end{document}